# Hybrid evolving clique-networks and their communicability

**Yimin Ding[1,2*], Bin Zhou[1], Xiaosong Chen[2]**

*[1]Faculty of Physics and Electronic Engineering, Hubei University, Wuhan 430062, China*

*[2]State Key Laboratory of Theoretical Physics, Institute of Theoretical Physics, Chinese Academy of Sciences, P.O. Box 2735, Beijing 100190, China*

*E-mail: dymhubu@sina.com

**Abstract** Aiming to understand real-world hierarchical networks whose degree distributions are neither power law nor exponential, we construct a hybrid clique network that includes both homogeneous and inhomogeneous parts, and introduce an inhomogeneity parameter to tune the ratio between the homogeneous part and the inhomogeneous one. We perform Monte-Carlo simulations to study various properties of such a network, including the degree distribution, the average shortest-path-length, the clustering coefficient, the clustering spectrum, and the communicability.

## 1. Introduction

The studies of complex networks have thrived over the past decades [1-4]. Various scholars have proposed many models to mimic the features of complex networks, including the Erdös-Rényi (ER) random-network model [5], the small-world-network model [6], and the scale-free-network model [7]. In most of these models, the networks are constructed from individual nodes [8-11]. In 2005, Derenyi *et. al.* introduced the concept of "clique" [12]. The concept of clique-percolation is later introduced by Palla *et al.* in the context of overlapping graph communities edges, and the clique percolation method is proposed to explore overlapping communities [13]. A clique of size $a$ is a completely interconnected unit that includes $a$ nodes and $a(a-1)$ /2 edges. Different from scale-free-networks, networks constructed from cliques possess hierarchical structures, and are able to offer deep insights into real-world networks whose central organizing principle is hierarchy [14-20].

Takemoto *et al* introduced a model studying evolving hierarchical networks constructed by reorganizing cliques according to the preferential attachment rule [21], and later several similar models are proposed. These models are applicable to inhomogeneous networks, which have power-law degree distributions [22-24]. However, many real-world networks are homogeneous. The homogeneity of a network means that almost all nodes of the network are topologically equivalent to one another, like those in regular lattices or in random graphs; and a homogeneous network usually has an exponential (or Poisson) degree distribution. To mimic the behaviors of these (homogeneous and hierarchical) networks, we constructed in ref. [25] a network by attaching cliques randomly, and the networks formed this way possess the desired properties, *viz.*, the networks are homogeneous and hierarchical.

In this paper, we study clique networks including both inhomogeneous and homogeneous parts. We introduce an *inhomogeneity parameter* $p$ which specifies the fraction of the homogeneous (or inhomogeneous) part in the whole network. By tuning the value of $p$, we are then able to change the network properties, in particular, the degree distribution. When $p$ decreases from 1 to 0, the degree distribution changes from a power-law distribution to an exponential one because a completely inhomogeneous network is changed to a homogeneous one. For $0<p<1$, our model can mimic the behaviors of real-world networks whose degree distributions are neither power law nor exponential. This paper is organized as follows. In Section 2, we describe how to construct our model network; in Section 3, we present the numerical results of the network properties; the study on the network communicability is reported in Section 4; a brief conclusion is presented in the last section.

## 2. Model

We consider a network composed of cliques of size $a$ ($a>2$). The cliques are linked to one another through common nodes and there is no extra edges added during the process of attaching (see Fig.1). At every time step, a new clique is attached to the network at $m$ ($m<a$) nodes. The selection rule of the attaching is specified as follows: (i) at a given probability $p$, the new clique selects the $m$ attachment nodes according to the preferential attachment (PA) rule, *viz*., the probability of node $i$ being attached by the new clique, $\Pi_i$, is given by

$$\Pi_i = \frac{k_i}{\sum_j k_j} \qquad (1)$$

where $k_i$ is the degree of node $i$ and the sum runs over all nodes of the network (excluding the new clique); (ii) at probability $1-p$, the new clique selects their attachment nodes in a completely random way, *viz*., $\Pi_i = 1/N(t)$, where $N(t)$ represents the total node number of the network at time step $t$ (excluding the new clique). Obviously, for $p=1$, our model is the same as the Takemoto-Oosawa model and becomes the scale-free model for $a=2$ and $m=1$; when $p=0$, this model is the same as the one reported in ref. 25 and becomes the complete random network model for $a=2$ and $m=1$. For a network with large $N$, the parameter $p$ specifies the fraction of the network that is inhomegeneous, and we thus call it *inhomogeneity parameter*.

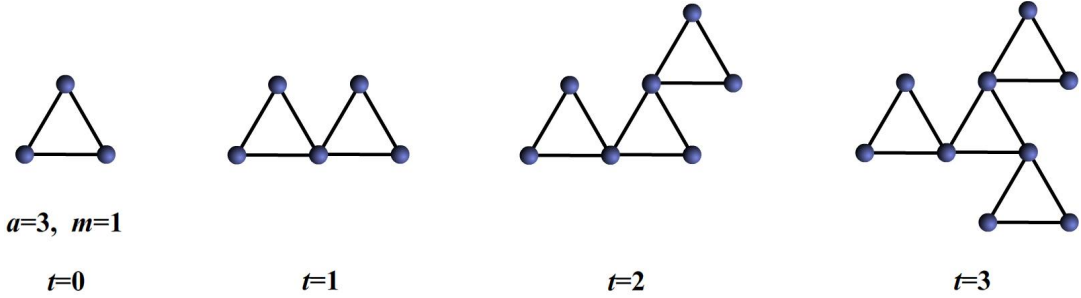

**Figure 1** (Color online). Schematic illustration of the evolutionary procedure of our model network for $a=3$ and $m=1$. Each clique is attached through common nodes with others, and no extra edge is added during the attaching process.

## 3. Network properties

We perform Monte Carlo simulations to study the properties of our model networks. We investigate (i) the degree distributions, (ii) the average shortest-path-lengths, (iii) the clustering coefficients, and (iv) the clustering spectra of networks having same node number and average degree but different inhomogeneity parameters. We focus on networks constructed from cliques with size $a=5$ and the number of attachment nodes $m=1$, 2 or 3. During the simulations, we tune the number of time steps to obtain networks with desired node numbers ($N$) and average degrees ($<k>$).

### 3.1. Degree distribution

The most important topological property of a complex network is its degree distribution $P(k)$ or cumulative degree distribution $CP(k)$. Considering that $CP(k)$ can reduce the noise in the tail of distribution curves, we present in Fig. 2 the numerical values of $CP(k)$ for $p=0$, 0.3, 0.5, 0.7, and 1, where we have set $N=5000$, $a=5$, $m=2$, and $<k>=20$. The inset of Figure 2 shows that the relation between $CP(k)$ and $k$ is represented by a straight line in a semi-log plot for $p=0$, implying that $CP(k)$ decays exponentially with $k$, which agrees with the results of a homogeneous network. For $p=1$, the relation between $CP(k)$ and $k$ in a log-log plot is a straight line whose slope is approximately -1.48, indicating that $CP(k) \propto k^{-\gamma}$, with $\gamma$=1.48, and that the degree distribution $P(k) \propto k^{-\gamma-1}$. The exponent of the power-law degree distribution is $2 < \gamma +1 < 3$, which agrees with the values reported elsewhere [1,3,4]. Furthermore, we can see from Figure 2 that when $p$ decreases from 1 to 0 the curves of $CP(k)$ v.s. $k$ in the log-log plots bend more and more (away from a straight line) because the homogeneous parts of the networks become larger and larger. In summary, we are able to change the degree distributions of the networks by tuning the value of $p$.

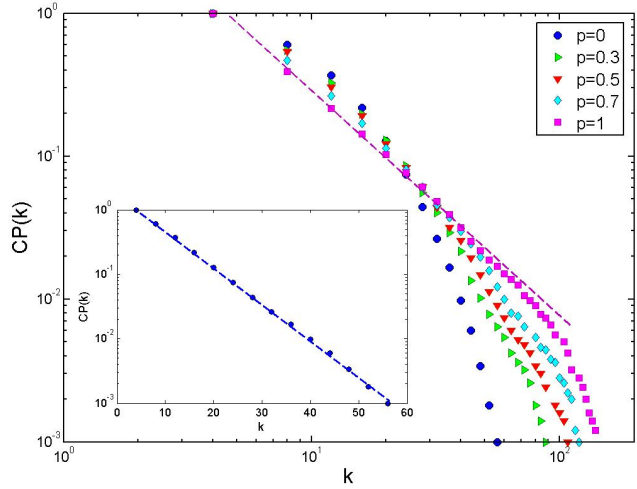

**Figure 2** (Color online). Cumulative degree distributions for $p$=0, 0.3, 0.5, 0.7, and 1, where the number of nodes $N$=5000, the clique size $a$=5, and the attachment-node number $m$=2.   Inserted is the CP distribution for $p$=0 re-plotted in a log-linear scale.

### 3.2. *Average shortest-path-length*

The average shortest-path-length ($L$) is another important property of complex networks. It is defined as the mean of the geodesic lengths ($d$) of all couples of nodes:

$$L = \frac{1}{N(N-1)} \sum_{i,j \in N, i \neq j} d_{ij} \qquad (2)$$

We consider a network with $N$=5000, $a$=5, and $m$=2. Because $m$ is decreased from 3 to 2, the average degree $<k>$ of the network is now much smaller (~6.66). We can see from Figure 3 that the average shortest-path-length $L$ decreases linearly with $p$. It is easy to understand the decrease of $L$ with $p$. Compared with the part of the network formed by attaching cliques in a random way, the part formed according to the PA rule is more clustered and thus has a smaller shortest-path-length; a larger $p$ means a larger fraction of the latter part and thus yields a smaller $L$. The reason behind the linearity between $L$ and $p$ is however not so transparent, and we will defer such discussion to our future work.

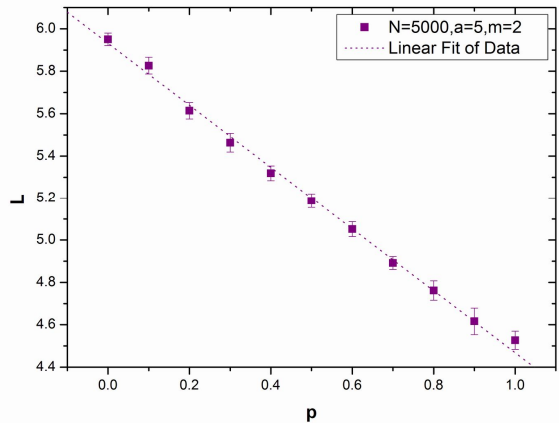

**Figure 3** (Color online). Relation between the average shortest-path-length $L$ and the inhomogeneity parameter $p$ for $N$=5000, $a$=5, $m$=2, and $<k>$=6.66.

### 3.3. Clustering coefficient

The clustering coefficient $c_i$ of node $i$ is defined as the ratio between the number of edges $e_i$ among the $k$ neighbors of node $i$ and the corresponding maximum possible value $k_i(k_i-1)/2$:

$$c_i = \frac{2e_i}{k_i(k_i-1)}. \tag{3}$$

The clustering coefficient $C$ of the whole network can then be obtained by averaging all $c_i$'s of the network:

$$C = \frac{1}{N}\sum_{i\in N} c_i \tag{4}$$

In Figure 4, we show the numerical values of $C$ for various $p$'s; we can see that it now increases linearly with $p$, because of a reason similar to that mentioned in Section 3.2.

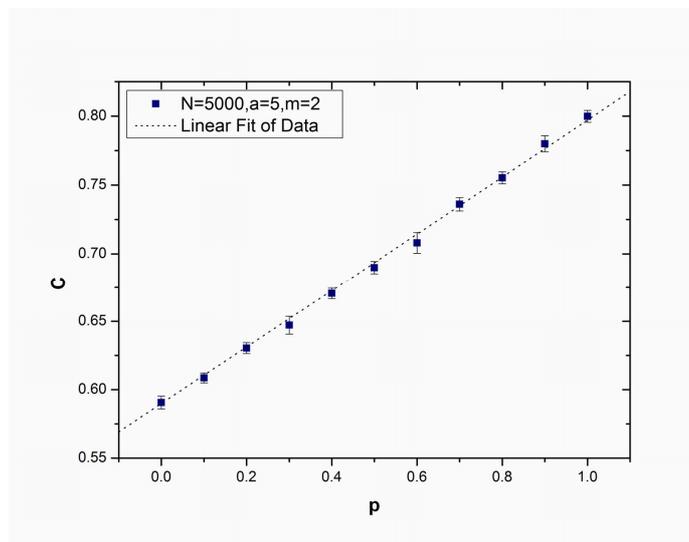

**Figure 4** (Color online). Relation between the clustering coefficient $C$ and the inhomogeneity parameter $p$ for $N$=5000, $a$=5, $m$=2, and $<k>$=6.66.

We also compare the clustering coefficient and average shortest-path-length of the ER random network with those of our model networks. The results are given in Table 1, which shows our networks have much larger clustering coefficients and smaller average shortest path lengths. Therefore, our model opens up a new way to construct small-world networks.

**Table 1**. Clustering coefficients and average shortest-path-lengths of networks with $a$=5, $m$=2, $N$=5000 and $<k>$=6.66.

| Network | ER random network | Our network ($p$=0) | Our network ($p$=0.5) | Our network ($p$=1) |
|---|---|---|---|---|
| $C$ | 0.00134 | 0.591 | 0.689 | 0.799 |
| $L$ | 5.96 | 5.95 | 5.19 | 4.46 |

### 3.4. Clustering spectrum

We now proceed to investigate the clustering spectrum of our network. A clustering spectrum, which reflects the modularity hierarchy of a network, is defined as

$$C(k) = \frac{1}{N\,P(k)} \sum_{i=1}^{N} c_i \times \delta_{k,k_i}, \qquad (5)$$

where $k_i$ and $c_i$ denote, respectively, the degree and clustering coefficient of node $i$, and $\delta_{k,ki}$ is Kronecker's delta function. The value of $C(k)$ gives the average of the clustering coefficients of all the nodes with degree $k$. Generally speaking, if $C(k) \propto k^{-\alpha}$, with $\alpha$ being about 1, the corresponding network has a hierarchical structure[26]. The clustering spectra of our networks (for $p$=0, 0.5, 1, and $m$=1, 2, 3) are shown in Figure 5. The approximate linear relations on the log-log plots indicate that these $C(k)$'s take a power law functional form and the power exponents (i.e., $\alpha$'s, which are equivalent to the slopes of the straight lines on the log-log plots) are approximately 1. We thus conclude that our model networks (composed of cliques) have hierarchical structures. This conclusion agrees with previous studies [21, 25-27].

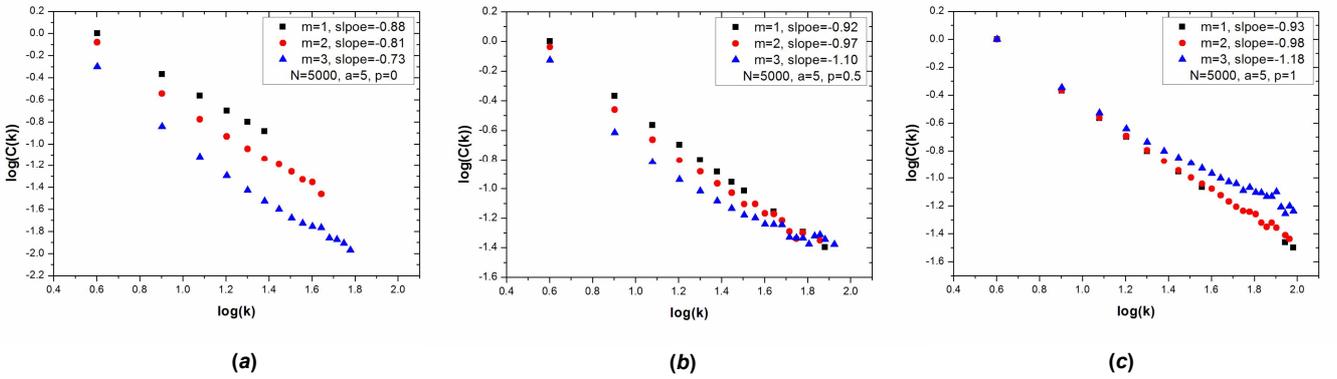

**Figure 5** (Color online). Clustering spectra of networks with $N$=5000, $a$=5, and $m$=1, 2, 3 for (a) $p$=0, (b) $p$=0.5, and (c) $p$=1.

## 4. Communicability

The network properties investigated in the previous section are limited to those related to individual nodes. Equally important is the communicability between different nodes, which provides quantitative measures of correlation between different parts of a system. Several definitions of communicability have been introduced and applied to a wide variety of real-world networks in recent years [28]. We follow the definition introduced in refs. [28, 29] and denote the communicability between nodes $u$ and $v$ as

$$G_{uv} = \sum_{n=0}^{\infty} \frac{(A^n)_{uv}}{n!} = (e^A)_{uv}, \qquad (6)$$

where $A$ is the adjacency matrix and its corresponding element is unity if the nodes $u$ and $v$ are linked to each other and is zero otherwise. The value of the $(u, v)$-entry of the $k$th power of the adjacency matrix, $(A^k)_{uv}$, thus gives the number of paths of length $k$ starting at the node $u$ and ending at the node $v$. In Equation (6) a factorial penalization (i.e., $1/n!$) has been assumed so that the shorter paths have more weight and longer ones have less. Note that the diagonal elements of the matrix $G_{uv}$ correspond to a weighted sum of the number of paths that originate from and end at same nodes. The Estrada index of a network is defined as the sum of all such diagonal elements, i.e., the trace of $G_{uv}$:

$$EE(G) = \mathrm{Tr}(e^A). \qquad (7)$$

The Estrada index is a global index for a network and commonly used to represent the communicability.

The numerical values of the Estrada indices of our networks for $p$=0, 0.5, 1 and $m$=1, 2, 3 is presented in Figure 6. For the purpose of comparison, the Estrada index of the ER network with the same node number and average degree is also shown. We can see that the Estrada indices of our networks are much larger than that of the ER network because of the complete connectivity of the cliques. Moreover, for our networks, the Estrada index is independent of *p* for *m*=1 and increases with *p* for *m*=2 and 3. Comparing the results for *m*=2 with those for *m*=3 yields that the larger *m* is the larger the influence of *p* on the Estrada index is. In addition, we can also see from Figure 6 that for *p*=0 the attachment number *m* has little influence on the Estrada index because we have set the average degrees of the networks with different *m*'s to be the same to one another; the Estrada index increases with *m* for nonzero *p*'s even though the corresponding networks have the same average degree.

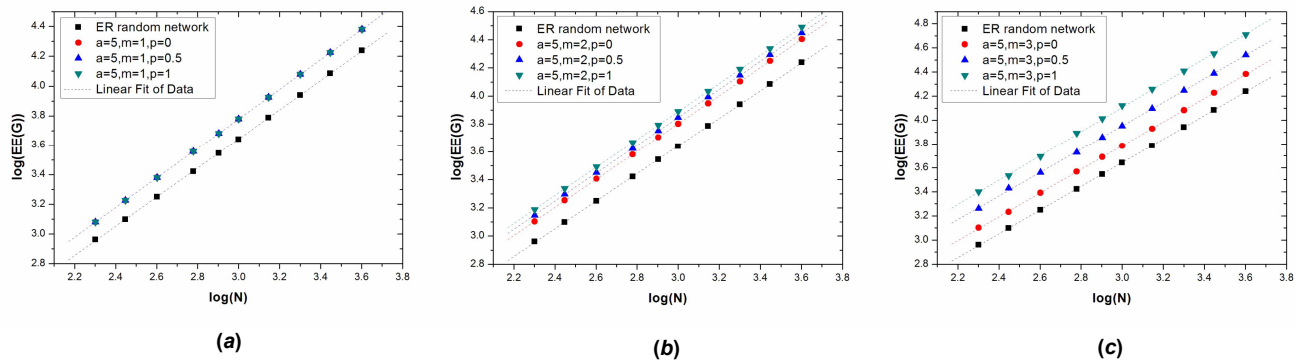

**Figure 6** (Color online). Relation between the Estrada index *EE(G)* and the network size *N* for *a*=5, *p*=0, 0.5, 1 and *m*=1, 2, 3.

## 5. Conclusion

With an aim to understand real-world hierarchical networks whose degree distributions are neither power law nor exponential, we have studied hybrid clique-networks including both homogeneous and inhomogeneous parts, and introduced an inhomogeneity parameter *p* to specify the fraction of the homogeneous (or inhomogeneous) parts in the networks. The numerical results show that our networks belong to the small-world networks. Their clustering coefficients are larger than that of the ER network and increase with *p*; and their average shortest-path-lengths are similar to that of the ER network and decrease with *p*. Our networks also exhibit a hierarchical modular structure. We have focused on numerical simulations; analytical studies will be explored in our future work.

## Acknowledgment

The authors thank Prof. Fangfu Ye for his valuable technical assistances and a careful revision of the manuscript, and thank Prof. Changping Yang for his help on the data analysis. This work was supported by the Open Foundation of the State Key Laboratory of Theoretical Physics in China (Grant No.Y3KF321CJ1), the National Natural Science Foundation of China (Grant No. 11274102), and the Program for New Century Excellent Talents in University of the Ministry of Education of China (Grant No. NCET-11-0960).

## References

[1]  R. Albert,  A.L. Barabási, Rev. Mod. Phys. 74 (2002) 47.

[2]  M. E. J. Newman, SIAM Rev. 45 (2003) 167.

[3]  S. Boccaletti, V. Latora, Y. Moreno, M. Chavez, D.-U. Hwang., Physics Reports 424 (2006) 175.

[4]  P. Holme, J.  Saramäki, Physics Reports 519 (2012) 97–125.

[5]  P. Erdös, A. Rényi, Publ. Math. Inst. Hung. Acad. Sci. 5 (1960)17.

[6]  D.J. Watts, S.H. Strogatz, Nature 393 (1998) 440.

[7]  A.L. Barabási,  R. Albert, Science 286 (1999) 509.

[8]  P. Sen *et al.*, Phys. Rev. E 67 (2003) 036106.

[9]  L.A.N.Amaral, A.Scala, M.Barthélémy, H.E.Stanley, Proc. Natl. Acad. Sci. U.S.A. 97 (2000) 11149.

[10] A.-L.Barabási, *et al.*, PhysicaA 311(2002) 590.

[11] B.Shargel, H.Sayama, I.R.Epstein, Y.Bar-Yam, Phys. Rev. Lett. 90 (2003) 068701.

[12] I.Derenyi, G.Palla, T.Vicsek, Phys.Rev.Lett. 94 (2005)160202.

[13] G.Palla, I.Derenyi, I.Farkas, T.Vicsek,Nature 435 (2005) 814.

[14] R. Milo *et al.*, Science 298 (2002) 824.

[15] Freeman, American Journal of Sociology, 98(1992) 152.

[16] Freeman, Social Networks, 18 (1996) 173.

[17] M.Girvanand, M.E.J.Newman, Proc. Natl. Acad. Sci. U.S.A. 99 (2002) 7821.

[18] Karrer, Newman, Phys.Rev.E 82 (2010) 066188.

[19] Evans J.Stat.Mech. (2010) P12037 (arXiv:1009.0638).

[20] A. Clauset, C. Moore, M. E. J. Newman, Nature 453 (2008) 98.

[21] K.Takemoto, C.Oosawa, Phys. Rev. E 72 (2005) 046116.

[22] K.Takemoto, C. Oosawa, T. Akutsua, Physica A 380 (2007) 665–672.

[23] L. Tan, J. Zhang, L. Jiang, J. Biol. Phys. 35 (2009) 197–207.

[24] T.S. Evans, J. Stat. Mech. (2010) P12037

[25] Y. Ding, B. Zhou, X. Chen, SCIENTIA SINICA Physica, Mechanica & Astronomica, 44 (2014) 299-304.

[26] E.Ravasz, A.-L.Barabási, Phys. Rev. E 67 (2003) 026112.

[27] Y. Ding, Z. Ding, C. Yang, Acta Phys. Sin.62 (2013) 098901.

[28] E.Estrada *et al*. Physics Reports 514 (2012) 89–119.

[29] E.R Estrada, N. Hatano, Phys.Rev.E 77 (2008) 036111.